\documentclass[sigconf]{acmart}

\AtBeginDocument{%
  }
\usepackage{multirow}
\usepackage{makecell}%
\usepackage{booktabs}

\begin{document}

\setcopyright{none}
\settopmatter{printacmref=false}
\renewcommand\footnotetextcopyrightpermission[1]{}

\title{Enhancing LLMs' Reasoning-Intensive Multimedia Search Capabilities through Fine-Tuning and Reinforcement Learning}

\author{Jinzheng Li$^1$, Sibo Ju$^2$, Yanzhou Su$^2$, Hongguang Li$^3$, Yiqing Shen$^{4,*}$}
\affiliation{%
  \institution{$^1$Fudan University\quad
  $^2$Fuzhou University\quad
  $^3$JF SmartInvest Holdings\quad
  $^4$ Johns Hopkins University\quad
  }
\city{}
\country{}
$^*$Corresponding author: yiqingshen1@gmail.com
}

\begin{abstract}
Existing large language models (LLMs) driven search agents typically rely on prompt engineering to decouple the user queries into search plans, limiting their effectiveness in complex scenarios requiring reasoning.
Furthermore, they suffer from excessive token consumption due to Python-based search plan representations and inadequate integration of multimedia elements for both input processing and response generation. 
To address these challenges, we introduce SearchExpert, a training method for LLMs to improve their multimedia search capabilities in response to complex search queries. 
Firstly, we reformulate the search plan in an efficient natural language representation to reduce token consumption. 
Then, we propose the supervised fine-tuning for searching (SFTS) to fine-tune LLM to adapt to these representations, together with an automated dataset construction pipeline. 
Secondly, to improve reasoning-intensive search capabilities, we propose the reinforcement learning from search feedback (RLSF) that takes the search results planned by LLM as the reward signals. 
Thirdly, we propose a multimedia understanding and generation agent that enables the fine-tuned LLM to process visual input and produce visual output during inference.
Finally, we establish an automated benchmark construction pipeline and a human evaluation framework.
Our resulting benchmark, SearchExpertBench-25, comprises 200 multiple-choice questions spanning financial and international news scenarios that require reasoning in searching. 
Experiments demonstrate that SearchExpert outperforms the commercial LLM search method (Perplexity Pro) by 36.60\% on the existing FinSearchBench-24 benchmark and 54.54\% on our proposed SearchExpertBench-25.
Human evaluations further confirm the superior readability.
\end{abstract}


\maketitle

\section{Introduction}
Large language models (LLMs) have transformed search interactions by enabling users to express complex natural language queries beyond simple keywords \cite{2407.00128,2307.03744,chatdev,li2023camel}. 
Specifically, existing agent-based search methods such as MindSearch~\cite{2407.20183}, FinSearch~\cite{2502.15684}, and MMSearch~\cite{2409.12959} leverage LLMs to decompose user queries into a directed acyclic graph (DAG) represented search plans, where nodes represent specific search keywords (and the corresponding search source), while edges denote logical dependencies between them.
However, MindSearch \cite{2407.20183} suffers from pre-determined search plans and temporal insensitivity.
Subsequently, FinSearch \cite{2502.15684} improved MindSearch with adaptive planning graphs and time-sensitive response generation for financial-related search queries, but remains limited in multimodal information processing \cite{2504.07957,2504.07942,2504.06606}.
Although MMSearch~\cite{2409.12959} incorporates visual input by converting them into text, it struggles to generate multimedia output.
Most importantly, all these existing agent-based search methods rely on prompt engineering to decompose queries \cite{2504.01523, 2503.09219}, which restricts their effectiveness in complex search scenarios involving reasoning.
Moreover, they also suffer from high excessive token consumption due to Python-based DAG representations, which limits practical efficiency \cite{2408.13885}.

\begin{table}[t!]
\centering
\caption{Comparison of LLM-driven search methods across major capabilities.
``\textbf{Efficient Representation}'' indicates whether the method employs token-efficient natural language representations for search plans rather than code-based DAG implementations.
``\textbf{SFT}'' shows if the method incorporates supervised fine-tuning specifically optimized for search planning.
``\textbf{RL}'' indicates whether the method leverages reinforcement learning to enhance searching or reasoning capabilities.
``\textbf{Multimedia Integration}'' refers to the ability to process visual inputs and generate visual elements in responses.
``\textbf{Complex Query Handling}'' demonstrates the capability to effectively process reasoning-intensive queries with multiple variables and deliberate obfuscation.
}
\label{tab:Comparison}
\renewcommand{\arraystretch}{1.2}
\resizebox{\columnwidth}{!}
{
\begin{tabular}{l|ccccc}
\hline
\textbf{Methods} &
\makecell[c]{\textbf{Efficient} \\ \textbf{Representation}} &
\makecell[c]{\textbf{SFT}} &
\makecell[c]{\textbf{RL}} &
\makecell[c]{\textbf{Multimedia} \\ \textbf{Integration}} &
\makecell[c]{\textbf{Complex Query} \\ \textbf{Handling}} \\
\hline
MindSearch~\cite{2407.20183} & \textcolor{red}{$\times$} & \textcolor{red}{$\times$} & \textcolor{red}{$\times$} & \textcolor{red}{$\times$} & \textcolor{red}{$\times$} \\
FinSearch~\cite{2502.15684} & \textcolor{red}{$\times$} & \textcolor{red}{$\times$} & \textcolor{red}{$\times$} & \textcolor{green}{\checkmark} & \textcolor{red}{$\times$} \\
MMSearch~\cite{2409.12959} & \textcolor{red}{$\times$} & \textcolor{red}{$\times$} & \textcolor{red}{$\times$} & \textcolor{green}{\checkmark} & \textcolor{red}{$\times$} \\
\hline
SearchExpert (Ours) & \textcolor{green}{\checkmark} & \textcolor{green}{\checkmark} & \textcolor{green}{\checkmark} & \textcolor{green}{\checkmark} & \textcolor{green}{\checkmark} \\
\hline
\end{tabular}
} 
\end{table}

To address these limitations, we introduce SearchExpert, a two-stage training framework that enhances LLMs' multimedia search capabilities through three following innovations.
First, we reformulate search plans using an efficient natural language representation for DAGs to reduce token consumption. 
Instead of Python code, we represent search graphs with text-based structures \cite{2502.10996,2502.10459} with supervised fine-tuning for searching (SFTS) to optimize LLMs specifically for generating effective search plans. 
Unlike general supervised fine-tuning (SFT) \cite{2011.01403,2409.15820} which requires manual data construction, we also introduce an automated pipeline to construct SFTS training data.
Second, to improve reasoning capabilities for complex search queries, we propose reinforcement learning from search feedback (RLSF).
Based on RLAIF \cite{2309.00267}, RLSF uses LLM to evaluate the quality of the search results and transforms these evaluations into reinforcement signals to update the LLM through Proximal Policy Optimization (PPO) \cite{1706.03741}. 
Our approach uniquely assesses both the semantic similarity to reference and the intrinsic quality of search results.
Third, we integrate a multimedia understanding and generation agent that enables SearchExpert to process visual inputs and generate visual outputs during inference, allowing for the handling of multimedia queries \cite{2503.01980}.
Table \ref{tab:Comparison} shows the high-level difference of our method from previous work.

Our major contributions are four-fold.
First, we propose an efficient natural language representation for search plans that replaces Python-based implementations with text-based DAG structures, reducing token consumption and processing time. 
Based on this representation, we propose supervised fine-tuning for searching (SFTS), a training method for LLMs in search contexts, along with an automated pipeline for constructing high-quality SFTS training datasets.
Second, we introduce reinforcement learning from search feedback (RLSF), which leverages the quality of search results as reward signals to improve the reasoning capabilities of LLM in complex queries. 
Unlike existing approaches that focus primarily on text quality, RLSF evaluates both semantic similarity to reference answers and intrinsic search result quality using LLM-based assessment.
Third, we propose a multimedia understanding and generation agent that enables our fine-tuned LLM to process visual input and produce visual output during inference. 
Fourth, we establish an automated benchmark construction methodology for complex search queries and present SearchExpertBench-25, a benchmark comprising 200 search questions spanning financial and international news domains that require reasoning in searching. 
We propose a human evaluation framework that assesses responses in multiple dimensions, including completeness, accuracy, timeliness, analytical integrity, and readability. 
Experimental results demonstrate that SearchExpert outperforms commercial and state-of-the-art methods.

\begin{figure*}[t!]
\centering
    \includegraphics[width=1\linewidth]{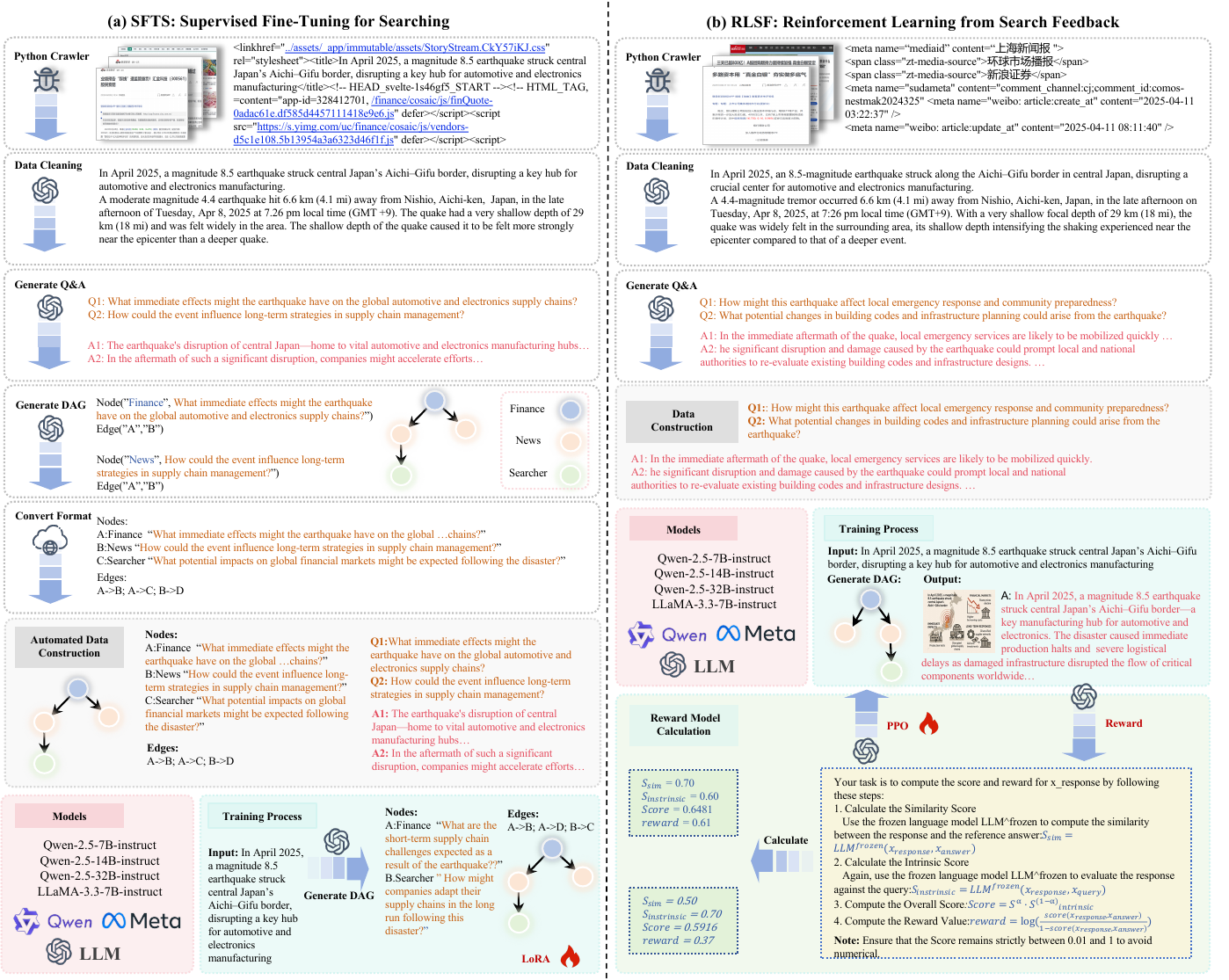}
        \caption{
The two-stage training framework of SearchExpert.
(a) Supervised fine-tuning for searching (SFTS): 
Our automated data construction pipeline begins with Python crawlers collecting recent online content, followed by data cleaning and Q\&A generation. 
We then generate efficient natural language DAG representations of search plans, converting from traditional Python-based formats to our token-efficient representation. 
The constructed training data is used to fine-tune LLMs to generate structured search plans directly from queries.
(b) Reinforcement searching from search feedback (RLSF): 
Building upon SFTS, our RLSF approach uses similar initial data collection but focuses on constructing more complex, reasoning-intensive search scenarios. 
The trained models generate search plans that are executed to retrieve information, with search quality evaluated through our dual-component reward mechanism combining semantic similarity and intrinsic quality assessment. 
The resulting scores guide LLM optimization through PPO.
}
    \label{fig:framework}
\end{figure*}

\section{Methods}
\subsection{Problem Formulation}
We formalize the reasoning-intensive multimedia search with LLMs as a process of generating search plans for complex queries.
Given a user query $q$ that contains both textual and visual elements, the LLM first generates a structured search plan $S$ represented as a directed acyclic graph (DAG) $G = (V, E)$, formalized as $G = \texttt{LLM}(q)$. 
The major focus of this work is to enhance the LLM's capability to generate effective search plans through our proposed SFTS and RLSF training, particularly for scenarios requiring complex reasoning.
In this formulation, $V = \{v_1, v_2, \ldots, v_n\}$ represents the set of search nodes, with each node containing specific keywords for the search along with the designated search source (\textit{e}.\textit{g}., general search, news search, financial search). 
The edge set $E \subseteq V \times V$ represents the logical dependencies between the search steps, where an edge $(v_i, v_j)$ indicates that the search results of node $v_i$ directly inform the query formulation for node $v_j$. 
To ensure a coherent and feasible search strategy, the graph must maintain acyclicity, preventing any sequence $v_{i_1} \rightarrow v_{i_2} \rightarrow \cdots \rightarrow v_{i_k} \rightarrow v_{i_1}$ from forming a cycle in $G$.
A major innovation in our approach is to represent $G$ in an efficient natural language format instead of the Python-based implementations used in previous work \cite{2407.20183}, which reduces token consumption and processing overhead.
Specifically, each node $v_i \in V$ is expressed as ``$v_i$: node feature,'' where the node feature comprises search keywords and the corresponding search source. 
The edges $E$ are represented as directional relationships between nodes, expressed as ``$v_i \rightarrow v_j$''. 
The complete representation can be formalized as follows:
\begin{equation}
\begin{aligned}
\text{Rep}(G) = \{(v_i: k_i) \mid v_i \in V, k_i \text{ is the keyword and search source}\} \\
\cup \{(v_i \rightarrow v_j) \mid (v_i, v_j) \in E\}.
\label{eq:dag}
\end{aligned}
\end{equation}
For example, a search plan investigating economic impacts might be represented as ``\textit{A: Japan earthquake impact (news search), B: global supply chain disruption (financial search), C: automotive industry response (industry search)... $A \rightarrow B$, $B \rightarrow C$}.'' 
This indicates a sequential search process where each node's results inform subsequent queries, enabling multi-step reasoning through structured information gathering.
After generation, the search plan $S$ is executed to retrieve relevant information, which is then synthesized into a final response $R$ by another frozen LLM. 
This two-stage formulation separates search planning from response synthesis, allowing us to focus our SFTS and RLSF training specifically on optimizing the search planning capabilities of the LLM, which directly impacts the quality and relevance of the final response.

\begin{figure*}[t!]
\centering
\includegraphics[width=1\linewidth]{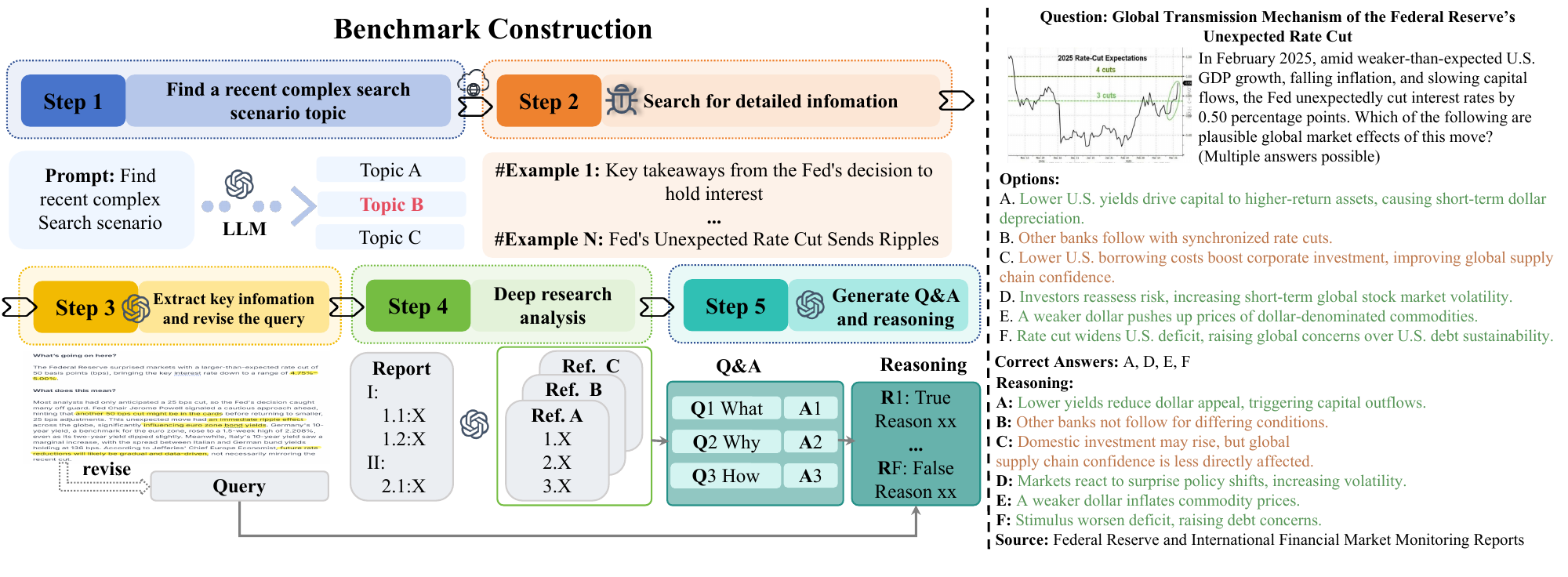}
    \caption{
Benchmark construction method and example case from SearchExpertBench-25.
The left panel illustrates our five-step pipeline for constructing reasoning-intensive search benchmarks.
It begins with LLM-based identification of recent complex topics requiring multi-hop search and causal inference, followed by information retrieval, key information extraction, OpenAI deep research analysis, and finally, question generation.
The right panel showcases an example benchmark question from SearchExpertBench-25 examining the global transmission mechanisms of the Federal Reserve's unexpected rate cut in February 2025.
This complex scenario features multiple correct answers (A, D, E, F) with explicit reasoning paths for each option, demonstrating how our benchmark evaluates a model's ability to generate effective search plans for navigating ambiguous, multi-variable financial scenarios.
}
    \label{fig:BenchMark Construction}
\end{figure*}

\subsection{Supervised Fine-Tuning for Searching}

The goal of our SFTS is to train LLMs to generate the efficient natural language DAG representation $\text{Rep}(G)$ defined in Eq.~\eqref{eq:dag} when presented with a complex search query $q$. 
Unlike standard supervised fine-tuning approaches that rely on manually constructed datasets and generic optimization objectives, SFTS leverages an automated data construction pipeline specifically designed for search planning, as shown in Fig.~\ref{fig:framework}.

\subsubsection{Automated Data Construction}
Our automated data construction pipeline reduces the manual effort typically required for SFT while ensuring high-quality training data for search-specific scenarios. 
It begins by crawling recent online texts $x_\text{online}$ from various time-sensitive sources, including news outlets, financial publications, and government websites.
Focusing on current information ensures that the corresponding content does not appear in the base LLM's training corpus, thereby necessitating search capabilities rather than knowledge retrieval.
For each crawled text, we employ a frozen LLM to generate query-answer pairs, namely $(x_\text{query},x_\text{answer}) = \text{LLM}^\text{frozen}(x_\text{online})$. 
To verify that these pairs require actual search capabilities rather than relying on information already present in the LLM's knowledge, we filter it by presenting each query $x_\text{query}$ to a separate LLM trying to answer it without access to the corresponding online context $x_\text{online}$ and analyze its response, \textit{i}.\textit{e}.

\begin{equation}
\text{isNovel}(x_\text{query}) = 
\begin{cases}
\text{True}, & \text{if } \text{Sim}(\texttt{LLM}^\text{frozen}(x_\text{query}), x_\text{answer}) < \tau \\
\text{False}, & \text{otherwise}
\end{cases}
\end{equation}
where $\text{Sim}(\cdot, \cdot)$ measures the semantic similarity determined by LLM, and $\tau$ represents a threshold that determines sufficient novelty. 
This filtering step again distinguishes our method from standard SFT approaches by ensuring that the training data specifically targets search-dependent queries rather than general knowledge queries.
Next, we feed each qualified query $x_\text{query}$ into FinSearch to generate both a search plan encoded in Python format $G_\text{Python}$ and the corresponding final response $x_\text{response}$. 
We take an additional step of validating response quality before converting the to $G_\text{Python}$ our efficient representation:
\begin{equation}
\text{isAligned}(x_\text{response}, x_\text{answer}) = \begin{cases}
\text{True}, & \text{if } \text{Sim}(x_\text{response}, x_\text{answer}) > \delta \\
\text{False}, & \text{otherwise}
\end{cases}
\end{equation}
where $\delta$ is the alignment threshold.
For aligned responses, we transform the Python-based search plan representation into our natural language representation described in Eq.~\eqref{eq:dag} through $\text{Rep}(G) = \text{LLM}^\text{frozen} (G_\text{Python})$ to reduce the number of tokens that represent $G$ while preserving the logical structure of the search plan. 
The final output comprises pairs of queries and their corresponding representations of the natural language search plan $x_\text{train} = \{$ ``input:'' $x_\text{query}$, ``output:'' $\text{Rep}(G)\}$.

\subsubsection{Training Process}
Using the automated generated dataset in the format $x_\text{train} = \{$ "input:" $x_\text{query}$, "output:" $\text{Rep}(G)\}$, we train the LLM to generate natural language DAG representations directly from queries, expressed as:
\begin{equation}
\text{Rep}(G)_\text{pred} = \text{LLM}_\theta(x_\text{query}),
\end{equation}
where $\text{LLM}_\theta$ represents our target LLM with parameters $\theta$, and $\text{Rep}(G)_\text{pred}$ is the predicted natural language representation of the search plan. 
The optimization objective employs a standard cross-entropy loss function that aligns the outputs with the reference natural language DAG representations.

\subsection{Reinforcement Learning from Search Feedback}

While SFTS provides a foundation for generating efficient search plans with LLM, complex reasoning-intensive queries often require capabilities beyond what supervised learning alone can achieve \cite{2401.04925, 2212.100071, 2301.12726, 2503.09516}.
Our RLSF addresses this limitation through a novel reward feedback mechanism that directly optimizes for search result quality rather than merely imitating reference search plans \cite{2502.18407}.

\subsubsection{Automated Dataset Construction}
We construct the RLSF dataset using methods similar to SFTS, but with heightened emphasis on complex reasoning scenarios that require multi-step search processes. 
Unlike SFTS which relies on single-source crawled content, RLSF dataset construction employs diverse and dynamic sources to generate more challenging time-sensitive queries, which ensures exposure to a broader range of reasoning patterns and information dependencies.
For RLSF training, we only need the query-answer pairs $(x_\text{query},x_\text{answer})$ rather than predetermined search plans, as optimization is driven by the quality of the search outcome rather than the adherence to the structure of the search plan.

\subsubsection{Training Process}
RLSF training begins with the LLM that generates a representation of a natural language search plan based on the input query: $\text{Rep}(G)_\text{pred} = \text{LLM}_\theta(x_{\text{query}})$. 
This search plan is executed to retrieve information and generate the corresponding response:
\begin{equation}
x_\text{response} = \texttt{execute}(\text{Rep}(G)_\text{pred}).
\end{equation}
The major innovation in our RLSF is the dual component reward calculation that captures both alignment with reference answers and intrinsic quality metrics. 
For semantic similarity assessment, we prompt an LLM to compare the generated response with the reference answer:
\begin{equation}
S_{\text{sim}}= \texttt{LLM}^\text{frozen}(x_\text{response}, x_\text{answer}),
\end{equation}
where LLM is prompt to analyze the factual consistency, the coverage of information, and the contextual relevance between the two texts, producing a normalized similarity score between 0 and 1.
This component ensures that the search plan leads to responses that capture the essential information needed to address the query.
Complementing this, our intrinsic quality assessment employs another LLM assessment that evaluates the completeness, precision, coherence, and clarity of the response independently of any reference.
\begin{equation}
S_{\text{intrinsic}} = \texttt{LLM}^\text{frozen}(x_\text{response}, x_{\text{query}}).
\end{equation}
These two factors are combined through a weighted geometric mean that balances fidelity to reference answers with intrinsic search quality:
\begin{equation}
\begin{aligned}
\text{score} =
S_{\text{sim}}^{\alpha} \cdot S_{\text{intrinsic}}^{(1-\alpha)} ,
\end{aligned}
\end{equation}
where $\alpha \in [0,1]$ controls the relative importance of similarity versus intrinsic quality.
%
%
The resulting score is transformed into a reinforcement learning signal using a log-odds transformation:
\begin{equation}
\text{reward}= \log\left(\frac{\text{score}(x_\text{response}, x_\text{answer})}{1 - \text{score}(x_\text{response}, x_\text{answer})}\right).
\end{equation}
This reward signal guides the optimization of LLM through PPO \cite{1706.03741}.

\subsection{Multimedia Understanding and Generation}
While SFTS and RLSF enhance the LLM's ability to generate effective search plans and reason across complex information sources, handling multimedia content requires additional operation for both input understanding and output generation. 
Our multimedia agent integrates visual processing and generation capabilities into the SearchExpert.
To process multimodal queries containing visual elements, we implement a visual understanding pipeline based on BLIP-2 \cite{2301.12597}, a vision-language model that converts visual information into textual representations. 
Given a query $q$ that contains both textual components $q_{\text{text}}$ and visual components $q_{\text{visual}}$, our agent converts visual elements into textual descriptions $q_{\text{visual\_desc}} = \texttt{BLIP-2}(q_{\text{visual}})$.
The resulting textual description is then merged with the original textual query to form an enriched query \textit{i}.\textit{e}., $q_{\text{enriched}} = \texttt{merge}(q_{\text{text}}, q_{\text{visual\_desc}})$.
This enriched query is processed by our fine-tuned LLM to generate the natural language search plan $\text{Rep}(G)$. 
To improve response quality and information comprehensiveness, we incorporate a graph generation that produces visual elements based on textual descriptions via DALLE-3 \cite{2212.09748}, represented as $v_{\text{output}} = \texttt{DALLE-3}(g_{\text{desc}})$, where $g_{\text{desc}}$ is a textual description of the desired graphical output derived from search results, and $v_{\text{output}}$ is the generated visual element. 

\subsection{Benchmark Dataset Construction}
To evaluate SearchExpert's reasoning-intensive multimedia search capabilities, we developed SearchExpertBench-25, a benchmark comprising 200 complex multiple-choice questions spanning financial and international news domains\cite{2311.05812,2504.01001,2306.09296}.
As illustrated in Fig.~\ref{fig:BenchMark Construction}, our benchmark construction can be automated by a five-step process.
Firstly, we employ LLMs to identify recent complex search scenario topics and generate the initial queries.
Then, based on these initial queries, we use web crawlers to extract detailed information from reliable sources, including major financial publications, news outlets, and industry reports.
Afterwards, we leverage LLM to extract key information from the retrieved HTML content and revise the initial query to increase the complexity.
We then used OpenAI deep research to generate the report for these revised queries, which contains hierarchical sections and explicit references to source materials. 
Finally, we generate question-answer pairs from the report with LLM by emphasizing the ``What,'' ``Why,'' and ``How'' dimensions of the analyzed information. 
The LLM generates explicit chains-of-thought reasoning paths that connect information sources to conclusions, creating questions that specifically test multi-step reasoning abilities, depicted as
\begin{equation}
(Q, A, CoT) = \texttt{LLM}^{\text{frozen}}(\text{SynthesizedInfo}, \text{ReasoningPrompt})
\end{equation}
where $Q$ represents the generated multi-choice question, $A$ is the correct answer (can be multiple correct answers), and $CoT$ is the chain-of-thought reasoning required to arrive at the answer from the available information sources.

\subsection{Human Evaluation Framework}
To assess SearchExpert's reasoning-intensive multimedia search capabilities, we also propose a human evaluation framework consisting of five essential factors based on FinSphere \cite{2501.12399}.
Our evaluation criteria were weighted to reflect their relative importance in reasoning-intensive search tasks: completeness and dimensionality of conclusions (30\%), accuracy and sufficiency of supporting evidence (15\%), data timeliness (15\%), analytical completeness and integrity (30\%), and readability and fluency (10\%), which prioritizes reasoning quality and comprehensive analysis while still accounting for factual accuracy and presentation quality.
Each criterion addresses a specific aspect of search performance. 
Completeness and dimensionality evaluate how thoroughly the LLM explores multiple perspectives and relevant dimensions of the query, particularly important for complex scenarios that require reasoning across diverse information sources. 
Accuracy and sufficiency assess whether the supporting evidence is factually correct and adequately substantiates the conclusions drawn, directly measuring the effectiveness of the search plan in retrieving relevant information.
Data timeliness examines whether the LLM appropriately prioritizes recent information when temporally relevant, a critical capability for financial and news domain searches. 
Analytical completeness and integrity evaluate the logical coherence and depth of reasoning demonstrated in connecting multiple information sources to reach conclusions, directly measuring the effectiveness of our RLSF approach in enhancing reasoning capabilities.
Finally, readability and fluency assess the clarity and effectiveness of communication, including the integration of multimedia elements into responses. 
This criterion specifically evaluates how well our multimedia understanding and generation agent enables the comprehensible presentation of complex search results.
Our evaluation protocol used a panel of domain experts who independently evaluated LLM output using a 5-point Likert scale for each criterion, with detailed rubrics ensuring the consistent application of standards. 
The assessors were blinded to the specific model generating each response to avoid bias. 
For each search query, the evaluators were provided with the original query, the LLM response, and access to authoritative reference sources for fact checking.

\begin{figure*}[t!]
\centering
    \includegraphics[width=\linewidth]{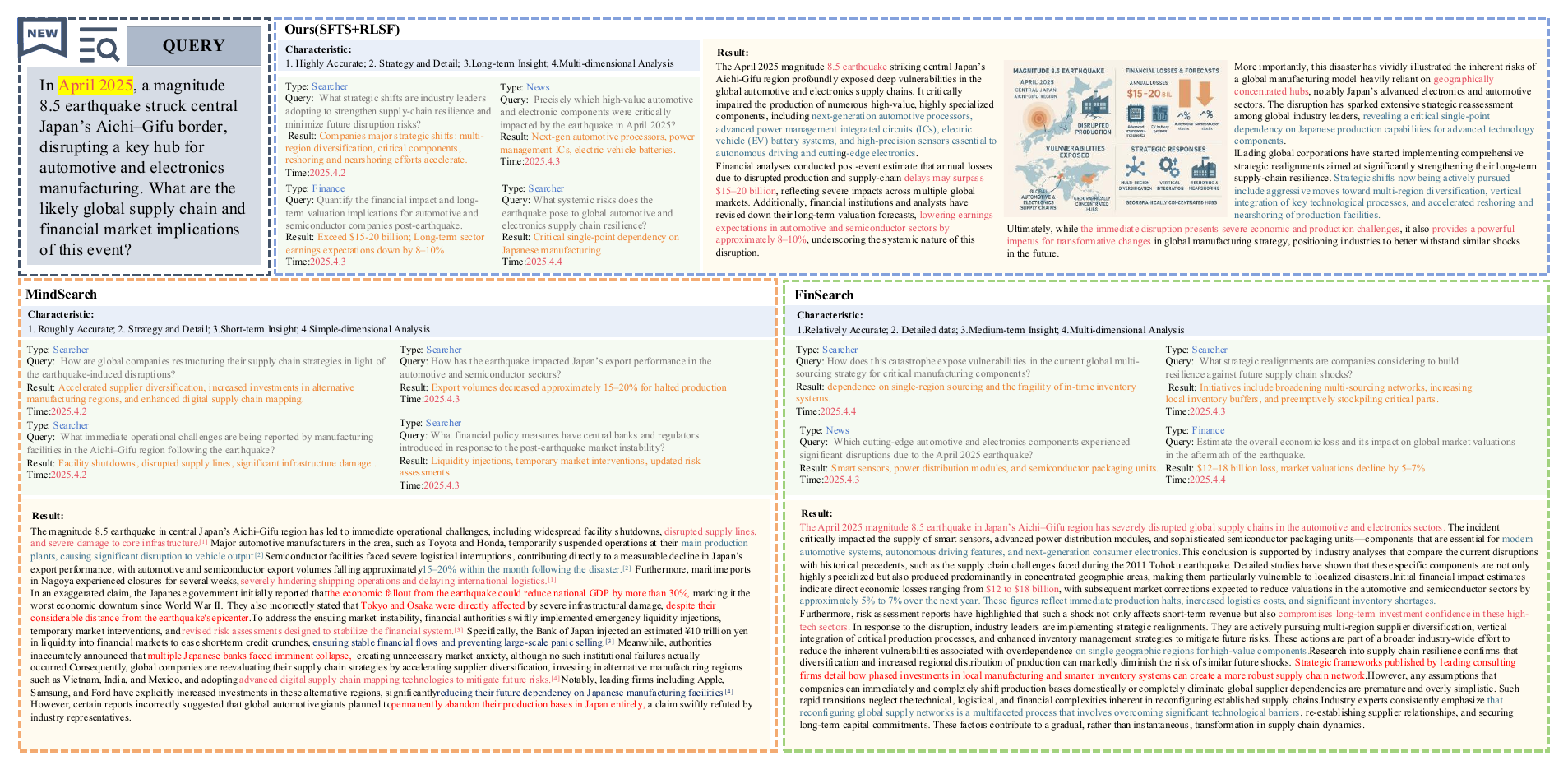}
       \caption{
Comparison of LLM-driven search methods analyzing a hypothetical April 2025 earthquake in Japan's manufacturing hub. 
Our SearchExpert (SFTS+RLSF) demonstrates superior reasoning capabilities through efficient natural language DAG representation, producing more comprehensive, long-term strategic analysis with precise component-level impacts and quantitative forecasts. 
MindSearch offers moderate but simpler dimensional analysis, while FinSearch provides detailed data but more limited causal reasoning. 
Green annotations indicate accurate reasoning; red shows analytical limitations; blue highlights distinctive information obtained through superior search planning.
}

    \label{fig:Case Study}
\end{figure*}

\section{Experiments}

\begin{table*}[t!]
\centering
\renewcommand{\arraystretch}{1.5}
\caption{Performance comparison of LLM-driven search methods on FinSearchBench-24 \cite{2502.15684} and SearchExpertBench-25, evaluating both effectiveness and efficiency. 
Our proposed SearchExpert method demonstrates accuracy improvements over existing approaches while maintaining reasonable computational efficiency through its efficient natural language representation of search plans. 
The accuracy metrics show the percentage of correct answers ($\pm$ standard deviation) across multiple models, demonstrating SearchExpert's consistent performance gains across model architectures. 
FinSearchBench-24 evaluates financial domain search capabilities, while SearchExpertBench-25 specifically tests complex reasoning-intensive multimedia search scenarios that require multi-hop information retrieval and causal inference.
Baseline refers to standalone LLMs without search capabilities, while SearchAgent \cite{2407.20183} and MindSearch \cite{2407.20183} represent general-purpose search frameworks with code-based DAG representations. 
FinSearch \cite{2502.15684} represents a financial-specialized search framework, and Perplexity Pro is a commercial AI search method.  
}
\setlength{\tabcolsep}{2.5pt} 
\label{tab:result1}
\resizebox{\textwidth}{!}{ 
\begin{tabular}{c|c |c c c| c c c}
\toprule
\multirow{2}{*}{Method} &\multirow{2}{*}{Models} & \multicolumn{3}{c|}{FinSearchBench-24 \cite{2502.15684}} & \multicolumn{3}{c}{SearchExpertBench-25} \\
    \cline{3-8}
    \multicolumn{1}{c|}{} & 
    \multicolumn{1}{c|}{} & 
    \multicolumn{1}{c}{\textbf{Accuracy (\%)}} & 
    \multicolumn{1}{c}{\textbf{Tokens (bit/answer)}} & 
    \multicolumn{1}{c|}{\textbf{Time (s/answer)}} &
    \multicolumn{1}{c}{\textbf{Accuracy (\%)}} &
    \multicolumn{1}{c}{\textbf{Tokens (bit/answer)}}&
    \multicolumn{1}{c}{\textbf{Time (s/answer)}} \\
     \hline
    \multirow{5}{*}{Baseline} &ChatGPT-4o \cite{2401.10744}  & $36.13 \pm \scriptstyle{1.22}$ & $293.82 \pm \scriptstyle{26.01}$ & $3.87 \pm \scriptstyle{0.15}$ & $22.50 \pm \scriptstyle{1.21}$ & $940.60 \pm \scriptstyle{45.98}$ & $6.04 \pm \scriptstyle{0.23}$ \\ 
     &Llama-3.3-7B-instruct \cite{2302.13971}  & $38.13 \pm \scriptstyle{1.27}$ & $\textbf{210.44} \pm \scriptstyle{23.64}$ & $4.96\pm \scriptstyle{0.25}$ & $21.00 \pm \scriptstyle{1.14}$ & $857.80 \pm \scriptstyle{55.28}$ & $7.14 \pm \scriptstyle{0.21}$ \\ 
     &Qwen2.5-7B-instruct \cite{2407.10671}  & $34.30 \pm \scriptstyle{1.12}$ & $255.39 \pm \scriptstyle{29.22}$ & $6.53 \pm \scriptstyle{0.21}$ & $21.50 \pm \scriptstyle{1.18}$ & $869.08 \pm \scriptstyle{66.19}$ & $6.83 \pm \scriptstyle{0.26}$ \\ 
     &Gemini-1.5 \cite{2312.11805}  & $35.47 \pm \scriptstyle{1.22}$ & $325.39 \pm \scriptstyle{39.61}$ & $5.04 \pm \scriptstyle{0.16}$ & $19.50 \pm \scriptstyle{1.35}$ & $926.05 \pm \scriptstyle{67.65}$ & $8.12 \pm \scriptstyle{0.31}$ \\ 
     &DeepSeek-v2.5  \cite{2405.04434}  & $34.07 \pm \scriptstyle{1.21}$  &$374.76 \pm \scriptstyle{33.49}$ & $14.50\pm \scriptstyle{0.27}$ & $23.00 \pm \scriptstyle{1.32}$ & $945.48 \pm \scriptstyle{61.67}$ & $7.81 \pm \scriptstyle{0.23}$ \\ 
    \hline
    \multirow{5}{*}{SearchAgent \cite{2407.20183}}  &ChatGPT-4o \cite{2401.10744} & $46.33\pm \scriptstyle{1.30}$ & $632.11\pm \scriptstyle{74.07}$ & $1.58\pm \scriptstyle{0.06}$ & $27.50\pm \scriptstyle{1.42}$ & $1649.86\pm \scriptstyle{143.18}$ & $10.31\pm \scriptstyle{0.16}$ \\ 
    &Llama-3.3-7B-instruct \cite{2302.13971}  & $43.80 \pm \scriptstyle{1.24}$ & $576.58 \pm \scriptstyle{42.67}$ & $1.61 \pm \scriptstyle{0.07}$    & $29.00 \pm \scriptstyle{1.31}$ & $1337.35 \pm \scriptstyle{122.07}$ & $11.42 \pm \scriptstyle{0.41}$ \\ 
    &Qwen2.5-7B-instruct \cite{2407.10671} & $41.21 \pm \scriptstyle{1.31}$ & $602.69 \pm \scriptstyle{56.55}$ & $1.51 \pm \scriptstyle{0.06}$ & $27.50 \pm \scriptstyle{1.74}$ & $1403.82 \pm \scriptstyle{132.87}$ & $11.81 \pm \scriptstyle{0.37}$ \\ 
    &Gemini-1.5 \cite{2312.11805} & $42.33 \pm \scriptstyle{1.28}$ & $654.85 \pm \scriptstyle{53.56}$ & $1.58 \pm \scriptstyle{0.04}$ & $31.50 \pm \scriptstyle{1.14}$ & $1586.78 \pm \scriptstyle{141.84}$ & $10.72 \pm \scriptstyle{0.36}$ \\ 
    &DeepSeek-v2.5 \cite{2405.04434} & $44.27 \pm \scriptstyle{1.26}$ & $697.92 \pm \scriptstyle{60.59}$ & $\textbf{1.32} \pm \scriptstyle{0.04}$ & $28.50 \pm \scriptstyle{1.52}$ & $1635.29 \pm \scriptstyle{158.10}$ & $12.10 \pm \scriptstyle{0.72}$ \\ 
    \hline
    \multirow{5}{*}{MindSearch \cite{2407.20183}}
     &ChatGPT-4o \cite{2401.10744} & $52.40 \pm \scriptstyle{1.33}$ & $3544.85 \pm \scriptstyle{473.24}$ & $19.09 \pm \scriptstyle{0.39}$ & $33.50 \pm \scriptstyle{1.12}$ & $4733.64 \pm \scriptstyle{356.78}$ & $24.12 \pm \scriptstyle{0.91}$ \\ 
    &Llama-3.3-7B-instruct \cite{2302.13971} & $53.60 \pm \scriptstyle{1.28}$ & $3098.40 \pm \scriptstyle{277.10}$ & $14.82 \pm \scriptstyle{0.45}$ & $35.50 \pm \scriptstyle{1.41}$ & $4352.94 \pm \scriptstyle{337.81}$ & $26.14 \pm \scriptstyle{0.34}$ \\ 
    &Qwen2.5-7B-instruct \cite{2407.10671} & $45.62 \pm \scriptstyle{1.26}$ & $3482.19 \pm \scriptstyle{298.21}$ & $18.12 \pm \scriptstyle{0.35}$ & $32.00 \pm \scriptstyle{1.17}$ & $4202.89 \pm \scriptstyle{316.71}$ & $23.42 \pm \scriptstyle{0.41}$ \\ 
    &Gemini-1.5 \cite{2312.11805} & $51.53 \pm \scriptstyle{1.29}$ & $3210.20 \pm \scriptstyle{223.32}$ & $20.14 \pm \scriptstyle{0.43}$ & $36.50 \pm \scriptstyle{0.97}$ & $4832.15 \pm \scriptstyle{377.39}$ & $25.72 \pm \scriptstyle{0.58}$ \\ 
    &DeepSeek-v2.5 \cite{2405.04434} & $49.73 \pm \scriptstyle{1.32}$  & $3741.20 \pm \scriptstyle{291.20}$ & $27.01 \pm \scriptstyle{0.58}$ & $33.00 \pm \scriptstyle{1.91}$ & $4899.24 \pm \scriptstyle{364.60}$ & $30.17 \pm \scriptstyle{0.93}$ \\ 
    \hline
     \multirow{5}{*}{FinSearch \cite{2502.15684}}
     &ChatGPT-4o \cite{2401.10744} & $76.20 \pm \scriptstyle{1.12}$ & $4828.21 \pm \scriptstyle{377.59}$ & $16.03 \pm \scriptstyle{0.43}$ & $44.50 \pm \scriptstyle{2.06}$ & $6734.26 \pm \scriptstyle{528.45}$ & $19.42 \pm \scriptstyle{0.81}$ \\ 
    &Llama-3.3-7B-instruct \cite{2302.13971} & $75.53 \pm \scriptstyle{1.04}$ & $4572.48 \pm \scriptstyle{393.61}$ & $14.55 \pm \scriptstyle{0.47}$ & $46.50 \pm \scriptstyle{1.81}$ & $7060.92 \pm \scriptstyle{644.80}$ & $21.31 \pm \scriptstyle{0.68}$ \\ 
    &Qwen2.5-7B-instruct \cite{2407.10671} & $75.40 \pm \scriptstyle{1.18}$ & $5482.19 \pm \scriptstyle{430.69}$ & $17.15 \pm \scriptstyle{0.45}$ & $41.00 \pm \scriptstyle{1.73}$ & $7120.88 \pm \scriptstyle{676.49}$ & $24.14 \pm \scriptstyle{0.78}$ \\ 
    &Gemini-1.5 \cite{2312.11805} & $74.87 \pm \scriptstyle{1.08}$  & $5954.77 \pm \scriptstyle{376.66}$ & $17.74  \pm \scriptstyle{0.53}$ & $47.50 \pm \scriptstyle{1.47}$ & $7368.72 \pm \scriptstyle{695.27}$ & $19.71 \pm \scriptstyle{0.91}$ \\ 
    &DeepSeek-v2.5 \cite{2405.04434} & $72.33 \pm \scriptstyle{1.15}$ & $5242.42 \pm \scriptstyle{458.70}$ & $29.31 \pm \scriptstyle{0.70}$ & $51.00 \pm \scriptstyle{2.42}$ & $7249.70 \pm \scriptstyle{706.33}$ & $20.73 \pm \scriptstyle{0.46}$ \\ 
    \hline
    Perplexity Pro   &sonar & $60.27 \pm \scriptstyle{1.26}$ & $\textbf{382.75} \pm \scriptstyle{29.66}$ & $5.85 \pm \scriptstyle{0.22}$ & $32.50 \pm \scriptstyle{1.74}$ & $\textbf{824.26} \pm \scriptstyle{68.22}$ & $\textbf{5.14} \pm \scriptstyle{0.84}$\\ 
    \hline
    \multirow{4}{*}{SearchExpert}
    &Llama-3.3-7B-instruct \cite{2302.13971}      & $69.98 \pm \scriptstyle{0.95}$ & $2156.20 \pm \scriptstyle{190.83}$ & $12.66 \pm \scriptstyle{0.48}$ & $64.00 \pm \scriptstyle{2.42}$ & $3681.29 \pm \scriptstyle{497.56}$ & $15.34 \pm \scriptstyle{0.42}$\\ 
    &Qwen2.5-7B-instruct  \cite{2407.10671} & $72.30 \pm \scriptstyle{0.57}$ & $2330.47 \pm \scriptstyle{196.89}$ & $13.48 \pm \scriptstyle{0.21}$ & $62.50 \pm \scriptstyle{1.82}$ & $3430.57 \pm \scriptstyle{424.59}$ & $14.85 \pm \scriptstyle{0.96}$\\ 
    &Qwen2.5-14B-instruct \cite{2407.10671}     & $76.82 \pm \scriptstyle{1.15}$ & $2480.26 \pm \scriptstyle{201.57}$ & $14.42 \pm \scriptstyle{0.88}$ & $67.00 \pm \scriptstyle{1.48}$ & $3717.43 \pm \scriptstyle{414.94}$ & $16.74 \pm \scriptstyle{0.72}$\\ 
    &Qwen2.5-32B-instruct \cite{2407.10671}     & $\textbf{82.33} \pm \scriptstyle{0.49}$ & $2813.62  \pm \scriptstyle{295.76}$ & $14.57 \pm \scriptstyle{0.21}$ & $\textbf{71.50} \pm \scriptstyle{1.37}$ & $4170.67 \pm \scriptstyle{427.09}$ & $18.72 \pm \scriptstyle{0.74}$ \\
    \bottomrule
    \end{tabular}
}
\end{table*}

\begin{table*}[t!]
  \centering
  \caption{
Ablation study evaluating the impact of our two-stage training approach on performance metrics across both benchmarks.
}
  \renewcommand{\arraystretch}{1.7}
  \label{tab:ablation}
  \resizebox{\textwidth}{!}{
  \begin{tabular}{c c| c c c| c c c}
   \toprule
    \multirow{2}{*}{\makecell[c]{SFTS\\Training}} &\multirow{2}{*}{\makecell[c]{RLSF\\Training}} &  \multicolumn{3}{c|}{FinSearchBench-24 \cite{2502.15684}} & \multicolumn{3}{c}{SearchExpertBench-25} \\
    \cline{3-8}
    \multicolumn{2}{c|}{} & 
    \multicolumn{1}{c}{\textbf{Accuracy (\%)}} & 
    \multicolumn{1}{c}{\textbf{Tokens (bit/answer)}} & 
    \multicolumn{1}{c|}{\textbf{Time (s/answer)}} &
    \multicolumn{1}{c}{\textbf{Accuracy (\%)}} &
    \multicolumn{1}{c}{\textbf{Tokens (bit/answer)}}&
    \multicolumn{1}{c}{\textbf{Time (s/answer)}}
    \\
    \hline
    $\times$ & $\times$       &$ 54.30 \pm \scriptstyle{\text{1.22}}$ &$ 3972.42 \pm \scriptstyle{\text{311.98}}$ & $\textbf{12.33} \pm \scriptstyle{\text{0.20}}$ & $39.50 \pm \scriptstyle{\text{1.74}}$ & $5864.20 \pm \scriptstyle{\text{674.52}}$ & $20.17 \pm \scriptstyle{0.39}$\\ 
    \hline
    \checkmark   & $\times$ & $70.30 \pm \scriptstyle{\text{1.39}}$ & $\textbf{2382.93} \pm \scriptstyle{\text{218.95}}$ & $15.57 \pm \scriptstyle{\text{0.21}}$ & $58.50 \pm \scriptstyle{\text{1.92}}$ & $\textbf{3937.68}   \pm \scriptstyle{\text{492.72}}$ & $\textbf{18.42} \pm \scriptstyle{0.37}$\\ \hline
    $\times$ &\checkmark  & $74.42 \pm \scriptstyle{\text{1.32}}$ & $4362.04   \pm \scriptstyle{\text{303.35}}$ & $14.83 \pm \scriptstyle{\text{1.19}}$ & $62.00 \pm \scriptstyle{\text{1.83}}$ & $6586.48 \pm \scriptstyle{\text{619.93}}$ & $19.32\pm \scriptstyle{0.13}$\\ \hline
    \checkmark &\checkmark  &$\textbf{82.33} \pm \scriptstyle{0.49}$ & $2813.62  \pm \scriptstyle{295.76}$ & $14.57 \pm \scriptstyle{0.21}$ & $\textbf{71.50} \pm \scriptstyle{1.37}$ & $4170.67 \pm \scriptstyle{427.09}$ & $23.22 \pm \scriptstyle{0.48}$\\ 
    \bottomrule
    \end{tabular}  
}
\end{table*}

\begin{figure}[htbp!]
\centering
    \includegraphics[width=1\linewidth]{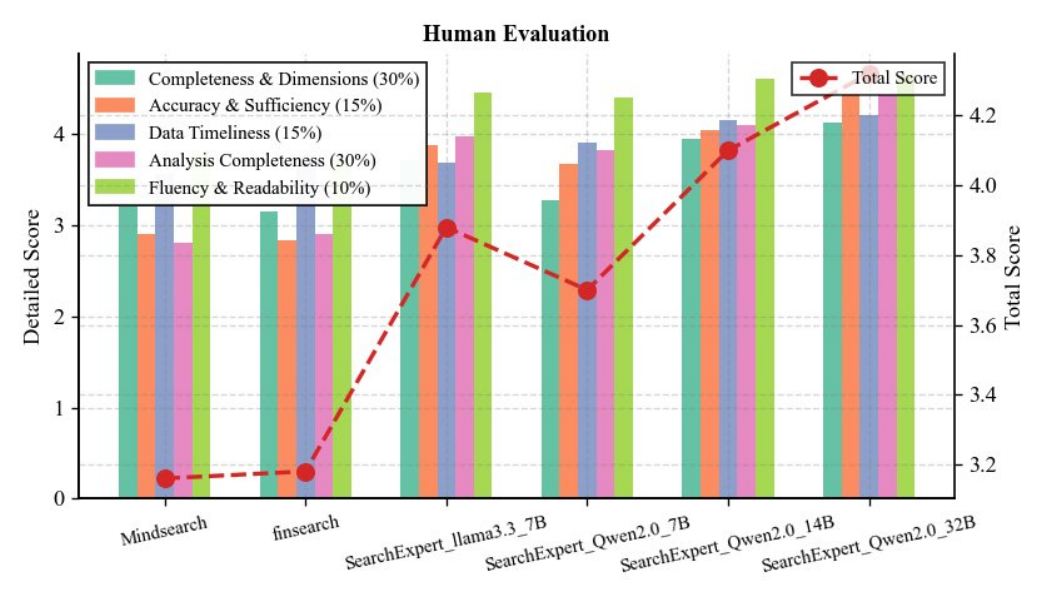}
        \caption{Human evaluation results comparing SearchExpert implementations across different model sizes against baseline search methods.
The total score (red dotted line) reveals that even smaller SearchExpert implementations outperform existing search frameworks, confirming that our two-stage training approach enhances reasoning-intensive search abilities while maintaining strong performance across all human evaluation dimensions.
}
    \label{fig:huamn}
\end{figure}

\subsection{Implementation Details}
We implemented SearchExpert using Python 3.10.4 in multiple LLM architectures to evaluate the effectiveness and generalizability of our approach.
Our experimental framework incorporated four state-of-the-art LLM backbones: 
Llama3.3-7B, Qwen-2.5-7B, Qwen-2.5-14B, and Qwen-2.5-32B \cite{2411.05934}, spanning different parameter scales to assess performance across model capacities.
For the SFTS stage, we leveraged LLaMA-Factory (v0.9.1) \cite{2403.13372}. 
For data construction, we use ChatGPT-4o (using both search and DeepResearch capabilities) and Perplexity. 
For the RLSF stage, we utilized trl, peft, and PyTorch 2.6.0 to allow for efficient fine-tuning. 
During this phase, we used FinSearch to execute the DAGs generated in natural language search and utilized ChatGPT-4o as an external reward model through API calls.

\subsection{Results}
Table \ref{tab:result1} presents the comparison of SearchExpert with state-of-the-art LLM-driven search methods on two benchmarks, namely FinSearchBench-24, which focuses on financial domain searches, and our newly constructed SearchExpertBench-25, which specifically targets multimedia search scenarios that require reasoning.
Our proposed two-stage training approach consistently outperforms existing methods on all model scales. 
In FinSearchBench-24, SearchExpert with Qwen2.5-32B-instruct achieves 82.33\% accuracy, surpassing the previously best-performing method, FinSearch (76.20\% with ChatGPT-4o). 
This improvement is even more pronounced on SearchExpertBench-25, where SearchExpert achieves 71.50\% accuracy, outperforming FinSearch with DeepSeek-v2.5 (51.00\%) by 20.50\% and Perplexity Pro (32.50\%) by 39.00\% points.
The performance gap between SearchExpert and other methods widens on SearchExpertBench-25, which contains complex queries. 
This validates our hypothesis that the combination of SFTS and RLSF enhances the LLM's ability to generate effective search plans for reasoning-intensive scenarios. 
Even with smaller models like Llama-3.3-7B-instruct and Qwen2.5-7B-instruct, SearchExpert achieves 64.00\% and 62.50\% accuracy respectively on SearchExpertBench-25, substantially outperforming all baseline and existing search methods.
In terms of computational efficiency, SearchExpert demonstrates a favorable balance between accuracy and efficiency. 
Although Perplexity Pro achieves the lowest token usage, namely 382.75 tokens per answer on FinSearchBench-24 and 824.26 on SearchExpertBench-25, its accuracy is substantially lower than SearchExpert. 
Compared to code-based DAG implementations used in MindSearch and FinSearch, our natural language representation of search plans reduces token consumption by approximately 42-53\%.
Processing time measurements show that SearchExpert maintains reasonable computational efficiency despite its improved reasoning capabilities. 
While not the fastest method (Perplexity Pro leads with 5.14 seconds per answer on SearchExpertBench-25), SearchExpert's processing times (14.85-18.72 seconds per answer) are substantially lower than those of MindSearch (23.42-30.17 seconds) and comparable to FinSearch (19.42-24.14 seconds).
Our results also reveal a clear correlation between model scale and performance within the SearchExpert framework. 
As the model size increases from 7B to 32B parameters, we observe consistent accuracy improvements on both benchmarks. 
This scaling trend suggests that larger models can better use our two-stage training approach to develop better search planning capabilities. 
As illustrated in Fig. \ref{fig:Case Study}, our case study further validates the effectiveness of our two-stage training approach.

\subsection{Human Evaluations}
As shown in Fig.~\ref{fig:huamn}, SearchExpert consistently outperforms existing search methods in all human evaluation criteria. 
MindSearch and FinSearch received substantially lower scores (approximately 3.2 on the 5-point scale) compared to even our smallest SearchExpert implementation (3.9 with Llama3.3-7B). 
Among the SearchExpert implementations, we observe a clear correlation between the model scale and human evaluation scores, with Qwen2.5-32B achieving the highest overall rating of 4.3. 
Particularly notable is the substantial improvement in analytical completeness scores, which increased from below 3.0 for baseline methods to over 4.0 for our larger models, confirming that RLSF effectively enhances reasoning capabilities by optimizing for search result quality. 
Furthermore, the high fluency and readability scores across all SearchExpert implementations demonstrate that our natural language representation of search plans maintains excellent presentation quality while reducing token consumption.

\subsection{Ablation Study} 
We conducted an ablation study using Qwen2.5-32B as the base model for both benchmarks in Table \ref{tab:ablation}.
The baseline model (without any training) achieves 54.30\% accuracy on FinSearchBench-24 and 39.50\% on SearchExpertBench-25, with relatively high token consumption (3972.42 and 5864.20 tokens per answer, respectively). 
This demonstrates the limitations of prompt-based search plan generation without specialized training.
When applying only SFTS, we observe a substantial improvement in both accuracy (70.30\% on FinSearchBench-24, 58.50\% on SearchExpertBench-25) and token efficiency (40.0\% reduction to 2382.93 tokens on FinSearchBench-24, 32.9\% reduction to 3937.68 tokens on SearchExpertBench-25), which validates our hypothesis that training LLMs to generate efficient natural language representations of search plans significantly reduces computational overhead while improving accuracy.
Conversely, applying only RLSF produces even higher accuracy gains (74.42\% on FinSearchBench-24, 62.00\% on SearchExpertBench-25) but at the cost of increased token consumption, which confirms that optimizing for search result quality through reinforcement learning substantially enhances the model's reasoning capabilities.
The complete SearchExpert achieves the highest accuracy across both benchmarks (82.33\% on FinSearchBench-24, 71.50\% on SearchExpertBench-25) while maintaining reasonable token efficiency (2813.62 and 4170.67 tokens per answer, respectively).
These results demonstrate the complementary nature of our two training components: SFTS provides token-efficient representations that reduce computational overhead, while RLSF enhances reasoning capabilities through optimization for search result quality. 
Their combination effectively addresses both the efficiency challenges and the reasoning challenges in existing LLM-driven search methods, particularly for complex scenarios.

\section{Conclusion}
In this paper, we introduce SearchExpert, a two-stage training framework that enhances LLMs' capabilities for reasoning-intensive search through complementary SFTS and RLSF. 
Our efficient natural language representation for search plans reduces token consumption compared to previous Python-based DAG implementations while maintaining readability. 
Experimental results demonstrate that SearchExpert substantially outperforms existing methods, surpassing commercial solutions such as Perplexity Pro by large margins. 
Human evaluations confirm SearchExpert's superior performance across all assessment dimensions, particularly in analytical completeness and reasoning depth, validating our approach's effectiveness for complex queries. 
Our multimedia agent further extends these capabilities by enabling visual input processing and output generation.
In general, the ability of SearchExpert to handle complex and reasoning-intensive search queries represents an advance for real-world applications.
One future direction can explore expanding SearchExpert to support cross-domain knowledge transfer and developing more advanced reasoning mechanisms for handling increasingly complex multi-hop search scenarios with uncertain or conflicting information sources. 
In addition, exploring multimodal understanding capabilities that can process and reason across video, audio, and other sensory inputs beyond current text and image modalities can be another future direction.

\bibliographystyle{ACM-Reference-Format}
\bibliography{ref}


\begin{thebibliography}{37}


\ifx \showCODEN    \undefined \def \showCODEN     #1{\unskip}     \fi
\ifx \showISBNx    \undefined \def \showISBNx     #1{\unskip}     \fi
\ifx \showISBNxiii \undefined \def \showISBNxiii  #1{\unskip}     \fi
\ifx \showISSN     \undefined \def \showISSN      #1{\unskip}     \fi
\ifx \showLCCN     \undefined \def \showLCCN      #1{\unskip}     \fi
\ifx \shownote     \undefined \def \shownote      #1{#1}          \fi
\ifx \showarticletitle \undefined \def \showarticletitle #1{#1}   \fi
\ifx \showURL      \undefined \def \showURL       {\relax}        \fi
\providecommand\bibfield[2]{#2}
\providecommand\bibinfo[2]{#2}
\providecommand\natexlab[1]{#1}
\providecommand\showeprint[2][]{arXiv:#2}

\bibitem[Caffagni et~al\mbox{.}(2025)]%
        {2503.01980}
\bibfield{author}{\bibinfo{person}{Davide Caffagni}, \bibinfo{person}{Sara Sarto}, \bibinfo{person}{Marcella Cornia}, \bibinfo{person}{Lorenzo Baraldi}, {and} \bibinfo{person}{Rita Cucchiara}.} \bibinfo{year}{2025}\natexlab{}.
\newblock \bibinfo{title}{Recurrence-Enhanced Vision-and-Language Transformers for Robust Multimodal Document Retrieval}.
\newblock
\showeprint{arXiv:2503.01980}


\bibitem[Cai and Jiang(2025)]%
        {2504.01523}
\bibfield{author}{\bibinfo{person}{Xuemeng Cai} {and} \bibinfo{person}{Lingxiao Jiang}.} \bibinfo{year}{2025}\natexlab{}.
\newblock \bibinfo{title}{Adapting Knowledge Prompt Tuning for Enhanced Automated Program Repair}.
\newblock
\showeprint{arXiv:2504.01523}


\bibitem[Catalano et~al\mbox{.}(2025)]%
        {2504.07942}
\bibfield{author}{\bibinfo{person}{Nico Catalano}, \bibinfo{person}{Stefano Samele}, \bibinfo{person}{Paolo Pertino}, {and} \bibinfo{person}{Matteo Matteucci}.} \bibinfo{year}{2025}\natexlab{}.
\newblock \bibinfo{title}{MARS: a Multimodal Alignment and Ranking System for Few-Shot Segmentation}.
\newblock
\showeprint{arXiv:2504.07942}


\bibitem[Chen et~al\mbox{.}(2024)]%
        {2407.20183}
\bibfield{author}{\bibinfo{person}{Zehui Chen}, \bibinfo{person}{Kuikun Liu}, \bibinfo{person}{Qiuchen Wang}, \bibinfo{person}{Jiangning Liu}, \bibinfo{person}{Wenwei Zhang}, \bibinfo{person}{Kai Chen}, {and} \bibinfo{person}{Feng Zhao}.} \bibinfo{year}{2024}\natexlab{}.
\newblock \bibinfo{title}{MindSearch: Mimicking Human Minds Elicits Deep AI Searcher}.
\newblock
\showeprint{arXiv:2407.20183}


\bibitem[Christiano et~al\mbox{.}(2017)]%
        {1706.03741}
\bibfield{author}{\bibinfo{person}{Paul Christiano}, \bibinfo{person}{Jan Leike}, \bibinfo{person}{Tom~B. Brown}, \bibinfo{person}{Miljan Martic}, \bibinfo{person}{Shane Legg}, {and} \bibinfo{person}{Dario Amodei}.} \bibinfo{year}{2017}\natexlab{}.
\newblock \bibinfo{title}{Deep reinforcement learning from human preferences}.
\newblock
\showeprint{arXiv:1706.03741}


\bibitem[de~Ocáriz~Borde et~al\mbox{.}(2024)]%
        {2408.13885}
\bibfield{author}{\bibinfo{person}{Haitz~Sáez de Ocáriz~Borde}, \bibinfo{person}{Anastasis Kratsios}, \bibinfo{person}{Marc~T. Law}, \bibinfo{person}{Xiaowen Dong}, {and} \bibinfo{person}{Michael Bronstein}.} \bibinfo{year}{2024}\natexlab{}.
\newblock \bibinfo{title}{Neural Spacetimes for DAG Representation Learning}.
\newblock
\showeprint{arXiv:2408.13885}


\bibitem[DeepSeek-AI et~al\mbox{.}(2024)]%
        {2405.04434}
\bibfield{author}{\bibinfo{person}{DeepSeek-AI}, \bibinfo{person}{Aixin Liu}, \bibinfo{person}{Bei Feng}, {et~al\mbox{.}}} \bibinfo{year}{2024}\natexlab{}.
\newblock \bibinfo{title}{DeepSeek-V2: A Strong, Economical, and Efficient Mixture-of-Experts Language Model}.
\newblock
\showeprint{arXiv:2405.04434}


\bibitem[Ding et~al\mbox{.}(2025)]%
        {2504.07957}
\bibfield{author}{\bibinfo{person}{Shengyuan Ding}, \bibinfo{person}{Shenxi Wu}, \bibinfo{person}{Xiangyu Zhao}, \bibinfo{person}{Yuhang Zang}, \bibinfo{person}{Haodong Duan}, \bibinfo{person}{Xiaoyi Dong}, \bibinfo{person}{Pan Zhang}, \bibinfo{person}{Yuhang Cao}, \bibinfo{person}{Dahua Lin}, {and} \bibinfo{person}{Jiaqi Wang}.} \bibinfo{year}{2025}\natexlab{}.
\newblock \bibinfo{title}{MM-IFEngine: Towards Multimodal Instruction Following}.
\newblock
\showeprint{arXiv:2504.07957}


\bibitem[Fu et~al\mbox{.}(2023)]%
        {2301.12726}
\bibfield{author}{\bibinfo{person}{Yao Fu}, \bibinfo{person}{Hao Peng}, \bibinfo{person}{Litu Ou}, \bibinfo{person}{Ashish Sabharwal}, {and} \bibinfo{person}{Tushar Khot}.} \bibinfo{year}{2023}\natexlab{}.
\newblock \bibinfo{title}{Specializing Smaller Language Models towards Multi-Step Reasoning}.
\newblock
\showeprint{arXiv:2301.12726}


\bibitem[Gao et~al\mbox{.}(2025a)]%
        {2504.06606}
\bibfield{author}{\bibinfo{person}{Minghe Gao}, \bibinfo{person}{Xuqi Liu}, \bibinfo{person}{Zhongqi Yue}, \bibinfo{person}{Yang Wu}, \bibinfo{person}{Shuang Chen}, \bibinfo{person}{Juncheng Li}, \bibinfo{person}{Siliang Tang}, \bibinfo{person}{Fei Wu}, \bibinfo{person}{Tat-Seng Chua}, {and} \bibinfo{person}{Yueting Zhuang}.} \bibinfo{year}{2025}\natexlab{a}.
\newblock \bibinfo{title}{Benchmarking Multimodal CoT Reward Model Stepwise by Visual Program}.
\newblock
\showeprint{arXiv:2504.06606}


\bibitem[Gao et~al\mbox{.}(2025b)]%
        {2502.10459}
\bibfield{author}{\bibinfo{person}{Yang Gao}, \bibinfo{person}{Hong Yang}, \bibinfo{person}{Yizhi Chen}, \bibinfo{person}{Junxian Wu}, \bibinfo{person}{Peng Zhang}, {and} \bibinfo{person}{Haishuai Wang}.} \bibinfo{year}{2025}\natexlab{b}.
\newblock \bibinfo{title}{LLM4GNAS: A Large Language Model Based Toolkit for Graph Neural Architecture Search}.
\newblock
\showeprint{arXiv:2502.10459}


\bibitem[Gunel et~al\mbox{.}(2020)]%
        {2011.01403}
\bibfield{author}{\bibinfo{person}{Beliz Gunel}, \bibinfo{person}{Jingfei Du}, \bibinfo{person}{Alexis Conneau}, {and} \bibinfo{person}{Ves Stoyanov}.} \bibinfo{year}{2020}\natexlab{}.
\newblock \bibinfo{title}{Supervised Contrastive Learning for Pre-trained Language Model Fine-tuning}.
\newblock
\showeprint{arXiv:2011.01403}


\bibitem[Han et~al\mbox{.}(2025)]%
        {2501.12399}
\bibfield{author}{\bibinfo{person}{Shijie Han}, \bibinfo{person}{Changhai Zhou}, \bibinfo{person}{Yiqing Shen}, \bibinfo{person}{Tianning Sun}, \bibinfo{person}{Yuhua Zhou}, \bibinfo{person}{Xiaoxia Wang}, \bibinfo{person}{Zhixiao Yang}, \bibinfo{person}{Jingshu Zhang}, {and} \bibinfo{person}{Hongguang Li}.} \bibinfo{year}{2025}\natexlab{}.
\newblock \bibinfo{title}{FinSphere: A Conversational Stock Analysis Agent Equipped with Quantitative Tools based on Real-Time Database}.
\newblock
\showeprint{arXiv:2501.12399}


\bibitem[Jiang et~al\mbox{.}(2024)]%
        {2409.12959}
\bibfield{author}{\bibinfo{person}{Dongzhi Jiang}, \bibinfo{person}{Renrui Zhang}, \bibinfo{person}{Ziyu Guo}, {et~al\mbox{.}}} \bibinfo{year}{2024}\natexlab{}.
\newblock \bibinfo{title}{MMSearch: Benchmarking the Potential of Large Models as Multi-modal Search Engines}.
\newblock
\showeprint{arXiv:2409.12959}


\bibitem[Jiang et~al\mbox{.}(2025)]%
        {2502.10996}
\bibfield{author}{\bibinfo{person}{Pengcheng Jiang}, \bibinfo{person}{Lang Cao}, \bibinfo{person}{Ruike Zhu}, \bibinfo{person}{Minhao Jiang}, \bibinfo{person}{Yunyi Zhang}, \bibinfo{person}{Jimeng Sun}, {and} \bibinfo{person}{Jiawei Han}.} \bibinfo{year}{2025}\natexlab{}.
\newblock \bibinfo{title}{RAS: Retrieval-And-Structuring for Knowledge-Intensive LLM Generation}.
\newblock
\showeprint{arXiv:2502.10996}


\bibitem[Jin et~al\mbox{.}(2025)]%
        {2503.09516}
\bibfield{author}{\bibinfo{person}{Bowen Jin}, \bibinfo{person}{Hansi Zeng}, \bibinfo{person}{Zhenrui Yue}, \bibinfo{person}{Jinsung Yoon}, \bibinfo{person}{Sercan Arik}, \bibinfo{person}{Dong Wang}, \bibinfo{person}{Hamed Zamani}, {and} \bibinfo{person}{Jiawei Han}.} \bibinfo{year}{2025}\natexlab{}.
\newblock \bibinfo{title}{Search-R1: Training LLMs to Reason and Leverage Search Engines with Reinforcement Learning}.
\newblock
\showeprint{arXiv:2503.09516}


\bibitem[Jin et~al\mbox{.}(2024)]%
        {2401.04925}
\bibfield{author}{\bibinfo{person}{Mingyu Jin}, \bibinfo{person}{Qinkai Yu}, \bibinfo{person}{Dong Shu}, \bibinfo{person}{Haiyan Zhao}, \bibinfo{person}{Wenyue Hua}, \bibinfo{person}{Yanda Meng}, \bibinfo{person}{Yongfeng Zhang}, {and} \bibinfo{person}{Mengnan Du}.} \bibinfo{year}{2024}\natexlab{}.
\newblock \bibinfo{title}{The Impact of Reasoning Step Length on Large Language Models}.
\newblock
\showeprint{arXiv:2401.04925}


\bibitem[Lee et~al\mbox{.}(2023)]%
        {2309.00267}
\bibfield{author}{\bibinfo{person}{Harrison Lee}, \bibinfo{person}{Samrat Phatale}, \bibinfo{person}{Hassan Mansoor}, \bibinfo{person}{Thomas Mesnard}, \bibinfo{person}{Johan Ferret}, \bibinfo{person}{Kellie Lu}, \bibinfo{person}{Colton Bishop}, \bibinfo{person}{Ethan Hall}, \bibinfo{person}{Victor Carbune}, \bibinfo{person}{Abhinav Rastogi}, {and} \bibinfo{person}{Sushant Prakash}.} \bibinfo{year}{2023}\natexlab{}.
\newblock \showarticletitle{RLAIF vs. RLHF: Scaling Reinforcement Learning from Human Feedback with AI Feedback}.
\newblock \bibinfo{howpublished}{Proceedings of the 41st International Conference on Machine Learning, PMLR 235:26874-26901, 2024}.
\newblock  (\bibinfo{year}{2023}).
\newblock
\showeprint{arXiv:2309.00267}


\bibitem[Lei et~al\mbox{.}(2023)]%
        {2311.05812}
\bibfield{author}{\bibinfo{person}{Yang Lei}, \bibinfo{person}{Jiangtong Li}, \bibinfo{person}{Dawei Cheng}, \bibinfo{person}{Zhijun Ding}, {and} \bibinfo{person}{Changjun Jiang}.} \bibinfo{year}{2023}\natexlab{}.
\newblock \bibinfo{title}{CFBenchmark: Chinese Financial Assistant Benchmark for Large Language Model}.
\newblock
\showeprint{arXiv:2311.05812}


\bibitem[Li et~al\mbox{.}(2023a)]%
        {li2023camel}
\bibfield{author}{\bibinfo{person}{Guohao Li}, \bibinfo{person}{Hasan Abed Al~Kader Hammoud}, \bibinfo{person}{Hani Itani}, \bibinfo{person}{Dmitrii Khizbullin}, {and} \bibinfo{person}{Bernard Ghanem}.} \bibinfo{year}{2023}\natexlab{a}.
\newblock \showarticletitle{CAMEL: Communicative Agents for "Mind" Exploration of Large Language Model Society}. In \bibinfo{booktitle}{\emph{Thirty-seventh Conference on Neural Information Processing Systems}}.
\newblock


\bibitem[Li et~al\mbox{.}(2023b)]%
        {2301.12597}
\bibfield{author}{\bibinfo{person}{Junnan Li}, \bibinfo{person}{Dongxu Li}, \bibinfo{person}{Silvio Savarese}, {and} \bibinfo{person}{Steven Hoi}.} \bibinfo{year}{2023}\natexlab{b}.
\newblock \bibinfo{title}{BLIP-2: Bootstrapping Language-Image Pre-training with Frozen Image Encoders and Large Language Models}.
\newblock
\showeprint{arXiv:2301.12597}


\bibitem[Li et~al\mbox{.}(2024)]%
        {2502.15684}
\bibfield{author}{\bibinfo{person}{Jinzheng Li}, \bibinfo{person}{Jingshu Zhang}, \bibinfo{person}{Hongguang Li}, {and} \bibinfo{person}{Yiqing Shen}.} \bibinfo{year}{2024}\natexlab{}.
\newblock \bibinfo{title}{An Agent Framework for Real-Time Financial Information Searching with Large Language Models}.
\newblock
\showeprint{arXiv:2502.15684}


\bibitem[Peebles and Xie(2022)]%
        {2212.09748}
\bibfield{author}{\bibinfo{person}{William Peebles} {and} \bibinfo{person}{Saining Xie}.} \bibinfo{year}{2022}\natexlab{}.
\newblock \bibinfo{title}{Scalable Diffusion Models with Transformers}.
\newblock
\showeprint{arXiv:2212.09748}


\bibitem[Pombal et~al\mbox{.}(2025)]%
        {2504.01001}
\bibfield{author}{\bibinfo{person}{José Pombal}, \bibinfo{person}{Nuno~M. Guerreiro}, \bibinfo{person}{Ricardo Rei}, {and} \bibinfo{person}{André F.~T. Martins}.} \bibinfo{year}{2025}\natexlab{}.
\newblock \bibinfo{title}{Zero-shot Benchmarking: A Framework for Flexible and Scalable Automatic Evaluation of Language Models}.
\newblock
\showeprint{arXiv:2504.01001}


\bibitem[Qian et~al\mbox{.}(2023)]%
        {chatdev}
\bibfield{author}{\bibinfo{person}{Chen Qian}, \bibinfo{person}{Wei Liu}, \bibinfo{person}{Hongzhang Liu}, \bibinfo{person}{Nuo Chen}, \bibinfo{person}{Yufan Dang}, \bibinfo{person}{Jiahao Li}, \bibinfo{person}{Cheng Yang}, \bibinfo{person}{Weize Chen}, \bibinfo{person}{Yusheng Su}, \bibinfo{person}{Xin Cong}, \bibinfo{person}{Juyuan Xu}, \bibinfo{person}{Dahai Li}, \bibinfo{person}{Zhiyuan Liu}, {and} \bibinfo{person}{Maosong Sun}.} \bibinfo{year}{2023}\natexlab{}.
\newblock \showarticletitle{ChatDev: Communicative Agents for Software Development}.
\newblock \bibinfo{journal}{\emph{arXiv preprint arXiv:2307.07924}} (\bibinfo{year}{2023}).
\newblock
\urldef\tempurl%
\url{https://arxiv.org/abs/2307.07924}
\showURL{%
\tempurl}


\bibitem[Spatharioti et~al\mbox{.}(2023)]%
        {2307.03744}
\bibfield{author}{\bibinfo{person}{Sofia~Eleni Spatharioti}, \bibinfo{person}{David~M. Rothschild}, \bibinfo{person}{Daniel~G. Goldstein}, {and} \bibinfo{person}{Jake~M. Hofman}.} \bibinfo{year}{2023}\natexlab{}.
\newblock \bibinfo{title}{Comparing Traditional and LLM-based Search for Consumer Choice: A Randomized Experiment}.
\newblock
\showeprint{arXiv:2307.03744}


\bibitem[Tahmid and Sarker(2024)]%
        {2411.05934}
\bibfield{author}{\bibinfo{person}{Saad Tahmid} {and} \bibinfo{person}{Sourav Sarker}.} \bibinfo{year}{2024}\natexlab{}.
\newblock \bibinfo{title}{Qwen2.5-32B: Leveraging Self-Consistent Tool-Integrated Reasoning for Bengali Mathematical Olympiad Problem Solving}.
\newblock
\showeprint{arXiv:2411.05934}


\bibitem[Team et~al\mbox{.}(2023)]%
        {2312.11805}
\bibfield{author}{\bibinfo{person}{Gemini Team}, \bibinfo{person}{Rohan Anil}, \bibinfo{person}{Sebastian Borgeaud}, {et~al\mbox{.}}} \bibinfo{year}{2023}\natexlab{}.
\newblock \bibinfo{title}{Gemini: A Family of Highly Capable Multimodal Models}.
\newblock
\showeprint{arXiv:2312.11805}


\bibitem[Touvron et~al\mbox{.}(2023)]%
        {2302.13971}
\bibfield{author}{\bibinfo{person}{Hugo Touvron}, \bibinfo{person}{Thibaut Lavril}, \bibinfo{person}{Gautier Izacard}, {et~al\mbox{.}}} \bibinfo{year}{2023}\natexlab{}.
\newblock \bibinfo{title}{LLaMA: Open and Efficient Foundation Language Models}.
\newblock
\showeprint{arXiv:2302.13971}


\bibitem[Xia et~al\mbox{.}(2025)]%
        {2502.18407}
\bibfield{author}{\bibinfo{person}{Yu Xia}, \bibinfo{person}{Jingru Fan}, \bibinfo{person}{Weize Chen}, \bibinfo{person}{Siyu Yan}, \bibinfo{person}{Xin Cong}, \bibinfo{person}{Zhong Zhang}, \bibinfo{person}{Yaxi Lu}, \bibinfo{person}{Yankai Lin}, \bibinfo{person}{Zhiyuan Liu}, {and} \bibinfo{person}{Maosong Sun}.} \bibinfo{year}{2025}\natexlab{}.
\newblock \bibinfo{title}{AgentRM: Enhancing Agent Generalization with Reward Modeling}.
\newblock
\showeprint{arXiv:2502.18407}


\bibitem[Xiong et~al\mbox{.}(2024)]%
        {2407.00128}
\bibfield{author}{\bibinfo{person}{Haoyi Xiong}, \bibinfo{person}{Jiang Bian}, \bibinfo{person}{Yuchen Li}, \bibinfo{person}{Xuhong Li}, \bibinfo{person}{Mengnan Du}, \bibinfo{person}{Shuaiqiang Wang}, \bibinfo{person}{Dawei Yin}, {and} \bibinfo{person}{Sumi Helal}.} \bibinfo{year}{2024}\natexlab{}.
\newblock \bibinfo{title}{When Search Engine Services meet Large Language Models: Visions and Challenges}.
\newblock
\showeprint{arXiv:2407.00128}


\bibitem[Yang et~al\mbox{.}(2024)]%
        {2407.10671}
\bibfield{author}{\bibinfo{person}{An Yang}, \bibinfo{person}{Baosong Yang}, \bibinfo{person}{Binyuan Hui}, {et~al\mbox{.}}} \bibinfo{year}{2024}\natexlab{}.
\newblock \bibinfo{title}{Qwen2 Technical Report}.
\newblock
\showeprint{arXiv:2407.10671}


\bibitem[Yang et~al\mbox{.}(2025)]%
        {2503.09219}
\bibfield{author}{\bibinfo{person}{Xinyi Yang}, \bibinfo{person}{Runzhe Zhan}, \bibinfo{person}{Derek~F. Wong}, \bibinfo{person}{Shu Yang}, \bibinfo{person}{Junchao Wu}, {and} \bibinfo{person}{Lidia~S. Chao}.} \bibinfo{year}{2025}\natexlab{}.
\newblock \bibinfo{title}{Rethinking Prompt-based Debiasing in Large Language Models}.
\newblock
\showeprint{arXiv:2503.09219}


\bibitem[Yu et~al\mbox{.}(2023)]%
        {2306.09296}
\bibfield{author}{\bibinfo{person}{Jifan Yu}, \bibinfo{person}{Xiaozhi Wang}, \bibinfo{person}{Shangqing Tu}, {et~al\mbox{.}}} \bibinfo{year}{2023}\natexlab{}.
\newblock \bibinfo{title}{KoLA: Carefully Benchmarking World Knowledge of Large Language Models}.
\newblock
\showeprint{arXiv:2306.09296}


\bibitem[Yuan et~al\mbox{.}(2024)]%
        {2401.10744}
\bibfield{author}{\bibinfo{person}{Ziqiang Yuan}, \bibinfo{person}{Kaiyuan Wang}, \bibinfo{person}{Shoutai Zhu}, \bibinfo{person}{Ye Yuan}, \bibinfo{person}{Jingya Zhou}, \bibinfo{person}{Yanlin Zhu}, {and} \bibinfo{person}{Wenqi Wei}.} \bibinfo{year}{2024}\natexlab{}.
\newblock \bibinfo{title}{FinLLMs: A Framework for Financial Reasoning Dataset Generation with Large Language Models}.
\newblock
\showeprint{arXiv:2401.10744}


\bibitem[Zhao et~al\mbox{.}(2024)]%
        {2409.15820}
\bibfield{author}{\bibinfo{person}{Yang Zhao}, \bibinfo{person}{Li Du}, \bibinfo{person}{Xiao Ding}, \bibinfo{person}{Kai Xiong}, \bibinfo{person}{Ting Liu}, {and} \bibinfo{person}{Bing Qin}.} \bibinfo{year}{2024}\natexlab{}.
\newblock \bibinfo{title}{Supervised Fine-Tuning Achieve Rapid Task Adaption Via Alternating Attention Head Activation Patterns}.
\newblock
\showeprint{arXiv:2409.15820}


\bibitem[Zheng et~al\mbox{.}(2024)]%
        {2403.13372}
\bibfield{author}{\bibinfo{person}{Yaowei Zheng}, \bibinfo{person}{Richong Zhang}, \bibinfo{person}{Junhao Zhang}, \bibinfo{person}{Yanhan Ye}, \bibinfo{person}{Zheyan Luo}, \bibinfo{person}{Zhangchi Feng}, {and} \bibinfo{person}{Yongqiang Ma}.} \bibinfo{year}{2024}\natexlab{}.
\newblock \bibinfo{title}{LlamaFactory: Unified Efficient Fine-Tuning of 100+ Language Models}.
\newblock
\showeprint{arXiv:2403.13372}


\end{thebibliography}

\end{document}